\documentclass{PoS}

\newcommand{\be}{\begin{equation}}
\newcommand{\eps}{\epsilon}
\newcommand{\e}{\epsilon}
\newcommand{\dps}{\displaystyle}
\newcommand{\ee}{\end{equation}}
\newcommand{\bea}{\begin{eqnarray}}
\newcommand{\eea}{\end{eqnarray}}
\newcommand{\loopint}[1]{\int \!\!\! \frac{d^D #1}{\left(2\pi\right)^D}\!}
\newcommand{\loopinttwo}[1]{\int \! \left[#1\right]}
\newcommand{\ESGamma}{S_{\Gamma}}

\newcommand{\lk}{\left(}
\newcommand{\rk}{\right)}
\newcommand{\lek}{\left[}
\newcommand{\rek}{\right]}
\newcommand{\lp}{\left.}
\newcommand{\rp}{\right.}
\newcommand{\nnb}{\nonumber}

\newcommand{\MB}[2]{\hs{-12} \int\limits_{\hs{15}_{ #1 -i \,
\infty}}^{\hs{15}^{ #1 +i\, \infty}} \hs{-15} \frac{d #2}{2\pi i}}
\newcommand{\pFq}[5]{\, \! _{#1} F_{#2}( #3 \, ; \, #4 \, ; \, #5 )}
\newcommand{\hs}[1]{\hspace*{#1 pt}}
\newcommand{\vs}[1]{\vspace*{#1 pt}}
\usepackage{cite}

\title{Master integrals for massless three-loop form factors}

\ShortTitle{Master integrals for massless three-loop form factors}

\author{\speaker{Tobias Huber} $^{a,b}$%
         \\
        \llap{$^a$}Institut f\"ur Theoretische Physik E, RWTH Aachen University,\\D-52056 Aachen, Germany\\
	\llap{$^b$}Fachbereich~7, Universit\"at~Siegen,\\ Walter-Flex-Str.~3, D-57068~Siegen, Germany\\
        E-mail: \email{huber@tp1.physik.uni-siegen.de}}

\abstract{We summarize the results for the master integrals of the three-loop quark and gluon form factor in massless QCD. Working in
dimensional regularization we extract poles up to $1/\eps^6$. The computational techniques involve, among others, the expansion of
higher transcendental functions and the Mellin-Barnes method. The coefficients of the Laurent expansion in $\eps$ are given either
analytically or numerically to high precision.}

\FullConference{RADCOR 2009 - 9th International Symposium on Radiative Corrections (Applications of Quantum Field Theory to Phenomenology) \\
		 October 25-30 2009\\
		 Ascona, Switzerland}

\begin{document}

\section{Introduction and computational methods}\label{sec:compmethods}

The quark form factor $\gamma^\ast \to q \bar q$ and gluon form factor
$H \to gg$ (effective coupling) are the simplest processes containing
IR divergences at higher orders in massless QFT,
and therefore are of particular interest in many aspects.
They have been used to predict the IR pole
structure of multi-leg amplitudes~\cite{Magnea:1990zb,catani,sterman,Dixon:2008gr}.
The form factors can also be exploited to extract resummation
coefficients~\cite{Magnea:2000ss,moch1}, and they enter the purely
virtual corrections to a number of collider reactions (Drell-Yan
process, Higgs production and decay, DIS).
Besides phenomenological applications, a major motivation for
obtaining analytic results at three-loop order and beyond is 
finding and understanding structures in massless gauge theories that
generalize to an arbitrary number of loops.  Much progress has
been achieved in the prediction of all-order singularity
structures in QCD~\cite{Becher:2009cu,Gardi:2009qi,Dixon:2009gx}, in conjectures
about the all-orders behaviour of maximally supersymmetric Yang-Mills
theories (see e.g.~\cite{Anastasiou:2003kj,Bern:2005iz,Bern:2006vw}) and in
investigations of the finiteness of N=8
supergravity, see e.g.~\cite{EarlyBernGravity1,EarlyBernGravity2,EarlyBernGravity3}.

The two-loop corrections to the massless-quark~\cite{vanneerven1,vanneerven2,vanneerven3} and
gluon~\cite{harlander,ravindran} form factors were computed in
dimensional regularisation with $D=4-2\e$ to order $\e^0$ and
subsequently extended to all orders in $\e$ in ref.~\cite{ghm}.
The three-loop form factors to order
$\e^{-1}$ (and $\e^0$ for contributions involving fermion loops in the quark form factor) were
computed in~\cite{moch1,moch2}. Recently, also the
three-loop form factors through $\e^0$ became available~\cite{BCSSS} (see also~\cite{Toedtli:2009bn}).

In order to calculate  the quark and gluon form factors
at higher orders in perturbation theory, the amplitudes
are reduced to a small set of master integrals by means of algebraic reduction
procedures~\cite{chet1,chet2,laporta,air,gr,fire,steinhausertalk}. 
At the three-loop level, the reduction results in 22 master integrals. Eight
of them are products of one-loop and two-loop vertex functions
or three-loop two-point functions, both of which are known
to sufficiently high orders in $\e$~\cite{ghm,chet1,chet2,bekavac,mincer}.
The remaining fourteen masters are genuine three-loop 
vertex functions which are depicted in Fig.~\ref{fig:3loopdiagrams}.
Each topology contains only one master 
integral, and corresponds to two-particle cuts of the
master integrals for massless four-loop off-shell propagator integrals~\cite{baikov}.
Working in dimensional regularisation with $D=4-2\eps$ and expanding the master
integrals in a Laurent series in $\e$, the finite part of the
three-loop form factors requires the extraction of all coefficients
through (polylogarithmic) weight six, i.e.\ coefficients containing terms up to
$\pi^6$ or $\zeta_3^2$.

The computational methods that we use during the calculation are manifold.
The easier integrals ($A_{5,1}$, $A_{5,2}$, $A_{6,1}$) contain only $\Gamma$-functions
and their expansion is straightforward. The more complicated masters
$A_{6,3}$, $A_{7,1}$, $A_{7,2}$, and $A_{7,4}$ are also represented in a
closed form in terms of hypergeometric functions of unit argument.
The latter are expanded in $\eps$ by means of {\tt HypExp}~\cite{hypexp,Huber:2007dx}.
The remaining integrals possess multiple Mellin-Barnes (MB)~\cite{smirnov,Tausk,smirnovbook,Alejandro,ambre}
representations. Their analytic continuation to $\eps=0$ was done with {\tt MB}~\cite{czakonMB}
and {\tt MBresolve}~\cite{S2MB}. These packages were also used for numerical cross
checks. In addition we performed numerical checks with the sector
decomposition methods of~\cite{gudrun1,gudrun2} and the {\tt FIESTA}~\cite{fiesta} package.

As analytic techniques we apply Barnes's lemmas and the theorem of
residues to the multiple MB integrals, and insert integral
representations of higher transcendental functions where appropriate. We also
make use of the {\tt HPL}~\cite{HPL}
and {\tt barnesroutines}~\cite{barnesroutines} packages,
as well as the nested sums algorithm~\cite{Vermaseren:1998uu,nestedsums}.

\section{Results}

\begin{figure}[p]
    \vs{20}
    \hs{31}
  \begin{tabular}{cccc}
  \parbox{3.3cm}{
\begin{picture}(0,0)
\thicklines
\put(-30,0){\vector(1,0){17}}
\put(-13,0){\line(1,0){13}}
\put(0,0){\line(3,2){40}}
\put(0,0){\line(3,-2){40}}
\put(0,0){\vector(3,2){37}}
\put(0,0){\vector(3,-2){37}}
\put(24,-16){\line(0,1){32}}
\qbezier[80](24,-15.9)(34,0)(24,16)
\qbezier[80](24,-16.1)(12.5,0)(24,16)
\end{picture}\\ \vs{30} \\ $A_{5,1}$}
  & 
\parbox{3.3cm}{
\begin{picture}(0,0)
\thicklines
\put(-30,0){\vector(1,0){17}}
\put(-13,0){\line(1,0){13}}
\put(0,0){\line(3,2){40}}
\put(0,0){\line(3,-2){40}}
\put(0,0){\vector(3,2){37}}
\put(0,0){\vector(3,-2){37}}
\put(24,-16){\line(0,1){32}}
\qbezier[80](24,-16)(34,0)(24,16)
\qbezier[80](0,0)(8,-20)(24,-16)
\end{picture}\\ \vs{30} \\ $A_{5,2}$}
  & 
\parbox{3.3cm}{
\begin{picture}(0,0)
\thicklines
\put(-30,0){\vector(1,0){17}}
\put(-13,0){\line(1,0){13}}
\put(0,0){\line(3,2){40}}
\put(0,0){\line(3,-2){40}}
\put(0,0){\vector(3,2){37}}
\put(0,0){\vector(3,-2){37}}
\qbezier[60](24,16.3)(18,12)(24,0)
\qbezier[60](24,-16.3)(18,-12)(24,0)
\qbezier[60](24,16.3)(30,12)(24,0)
\qbezier[60](24,-16.3)(30,-12)(24,0)
\end{picture}\\ \vs{30} \\ $A_{6,1}$}
  &
\parbox{3.3cm}{
\begin{picture}(0,0)
\thicklines
\put(-30,0){\vector(1,0){17}}
\put(-13,0){\line(1,0){13}}
\put(0,0){\line(3,2){40}}
\put(0,0){\line(3,-2){40}}
\put(0,0){\vector(3,2){37}}
\put(0,0){\vector(3,-2){37}}
\put(27,-18){\line(0,1){36}}
\put(26.8,-17.8){\line(-1,3){5.9}}
\put(26.8,17.8){\line(-1,-3){5.9}}
\put(0,0){\line(1,0){20.8}}
\end{picture}\\ \vs{30} \\ $A_{6,2}$} \\ \\ \\ \\ \\
  \end{tabular}
  \hs{50}
  \begin{tabular}{ccc}
  \parbox{4.5cm}{
\begin{picture}(0,0)
\thicklines
\put(-50,0){\vector(1,0){27}}
\put(-23,0){\line(1,0){23}}
\put(0,0){\line(3,2){60}}
\put(0,0){\line(3,-2){60}}
\put(0,0){\vector(3,2){57}}
\put(0,0){\vector(3,-2){57}}
\put(46.2,-30.8){\line(0,1){61.6}}
\put(18,12){\line(2,-3){28.7}}
\qbezier[80](17.6,12.9)(20,40)(46,31)
\end{picture}\\ \vs{40} \\ $A_{6,3}$}
 &
\parbox{4.5cm}{
\begin{picture}(0,0)
\thicklines
\put(-50,0){\vector(1,0){27}}
\put(-23,0){\line(1,0){23}}
\put(0,0){\line(3,2){60}}
\put(0,0){\line(3,-2){60}}
\put(0,0){\vector(3,2){55}}
\put(0,0){\vector(3,-2){55}}
\put(24,-16){\line(2,5){17.4}}
\put(42,-28){\line(-1,2){10.5}}
\put(27,2){\line(-1,2){6}}
\qbezier[60](0,0)(0,26)(20.7,14)
\end{picture}\\ \vs{40} \\ $A_{7,1}$}
  &   
  \parbox{4.5cm}{
\begin{picture}(0,0)
\thicklines
\put(-50,0){\vector(1,0){27}}
\put(-23,0){\line(1,0){23}}
\put(0,0){\line(3,2){60}}
\put(0,0){\line(3,-2){60}}
\put(0,0){\vector(3,2){55}}
\put(0,0){\vector(3,-2){55}}
\put(24,-16){\line(2,5){17.4}}
\put(42,-28){\line(-1,2){10.5}}
\put(27,2){\line(-1,2){6}}
\qbezier[60](20.7,14)(20,40)(41,27.7)
\end{picture}\\ \vs{40} \\ $A_{7,2}$}    \\ \\ \\ \\ \\
\parbox{4.5cm}{
\begin{picture}(0,0)
\thicklines
\put(-50,0){\vector(1,0){27}}
\put(-23,0){\line(1,0){23}}
\put(0,0){\line(3,2){60}}
\put(0,0){\line(3,-2){60}}
\put(0,0){\vector(3,2){57}}
\put(0,0){\vector(3,-2){57}}
\put(46.2,-30.8){\line(0,1){61.6}}
\put(18,12){\line(2,-3){28.7}}
\put(18,-12){\line(0,1){24}}
\end{picture}\\ \vs{40} \\ $A_{7,3}$}
 &
 \parbox{4.5cm}{
\begin{picture}(0,0)
\thicklines
\put(-50,0){\vector(1,0){27}}
\put(-23,0){\line(1,0){23}}
\put(0,0){\line(3,2){60}}
\put(0,0){\line(3,-2){60}}
\put(0,0){\vector(3,2){57}}
\put(0,0){\vector(3,-2){57}}
\put(18,-12){\line(0,1){24}}
\put(18,-12){\line(2,3){6.7}}
\put(28,3){\line(2,3){18.5}}
\put(18,12){\line(2,-3){28.7}}
\end{picture}\\ \vs{40} \\ $A_{7,4}$}
&
\parbox{4.5cm}{
\begin{picture}(0,0)
\thicklines
\put(-50,0){\vector(1,0){27}}
\put(-23,0){\line(1,0){23}}
\put(0,0){\line(3,2){60}}
\put(0,0){\line(3,-2){60}}
\put(0,0){\vector(3,2){57}}
\put(0,0){\vector(3,-2){57}}
\put(46.2,-30.8){\line(0,1){61.6}}
\put(18,-12){\line(2,3){6.7}}
\put(28,3){\line(2,3){18.5}}
\put(18,12){\line(2,-3){28.7}}
\end{picture}\\ \vs{40} \\ $A_{7,5}$} \\ \\ \\ \\ \\
 \parbox{4.5cm}{
\hs{45}
\begin{picture}(0,0)
\thicklines
\put(-50,0){\vector(1,0){27}}
\put(-23,0){\line(1,0){23}}
\put(0,0){\line(3,2){64}}
\put(0,0){\line(3,-2){64}}
\put(0,0){\vector(3,2){60}}
\put(0,0){\vector(3,-2){60}}
\put(30,-20){\line(0,1){40}}
\put(30,-20){\line(-4,5){20.7}}
\put(18,-5){\line(5,6){9}}
\put(34,14.2){\line(5,6){15.5}}
\end{picture}\\ \vs{40} \\\hs{35} $A_{8}$}
  &   \parbox{4.5cm}{
\hs{60}
\begin{picture}(0,0)
\thicklines
\put(-50,0){\vector(1,0){27}}
\put(-23,0){\line(1,0){23}}
\put(0,0){\line(3,2){60}}
\put(0,0){\line(3,-2){60}}
\put(0,0){\vector(3,2){55}}
\put(0,0){\vector(3,-2){55}}
\put(24,-16){\line(0,1){32}}
\put(42,-28){\line(0,1){56}}
\put(24,0){\line(1,0){18}}
\end{picture}\\ \vs{40} \\ \hs{55} $A_{9,1}$} \\ \\ \\ \\ \\
\parbox{4.5cm}{
\hs{45}
\begin{picture}(0,0)
\thicklines
\put(-50,0){\vector(1,0){27}}
\put(-23,0){\line(1,0){23}}
\put(0,0){\line(3,2){64}}
\put(0,0){\line(3,-2){64}}
\put(0,0){\vector(3,2){60}}
\put(0,0){\vector(3,-2){60}}
\put(18,-12){\line(0,1){24}}
\put(30,-20){\line(0,1){40}}
\put(18,-5){\line(5,6){9}}
\put(34,14.2){\line(5,6){15.5}}
\end{picture}\\ \vs{40} \\ \hs{55} $A_{9,2}$}
&
 \parbox{4.5cm}{
\hs{60}
\begin{picture}(0,0)
\thicklines
\put(-50,0){\vector(1,0){27}}
\put(-23,0){\line(1,0){23}}
\put(0,0){\line(3,2){60}}
\put(0,0){\line(3,-2){60}}
\put(0,0){\vector(3,2){56}}
\put(0,0){\vector(3,-2){56}}
\put(24,-16){\line(0,1){13.4}}
\put(42,-28){\line(0,1){25.6}}
\put(23.8,-2.6){\line(1,0){18.4}}
\put(42,-2.5){\line(-2,3){14}}
\put(24.1,-2.6){\line(2,3){7}}
\put(35.1,13.9){\line(2,3){11.3}}
\end{picture}\\ \vs{40} \\ \hs{55} $A_{9,4}$} \\ \\ \\
  \end{tabular}
    \vs{-30}
  \caption{The fourteen genuine vertex-type master integrals. The incoming momentum is $q=p_1+p_2$. 
  Outgoing momenta are taken to be on-shell and massless, $p_1^2=p_2^2=0$.  \label{fig:3loopdiagrams}}
  \end{figure}
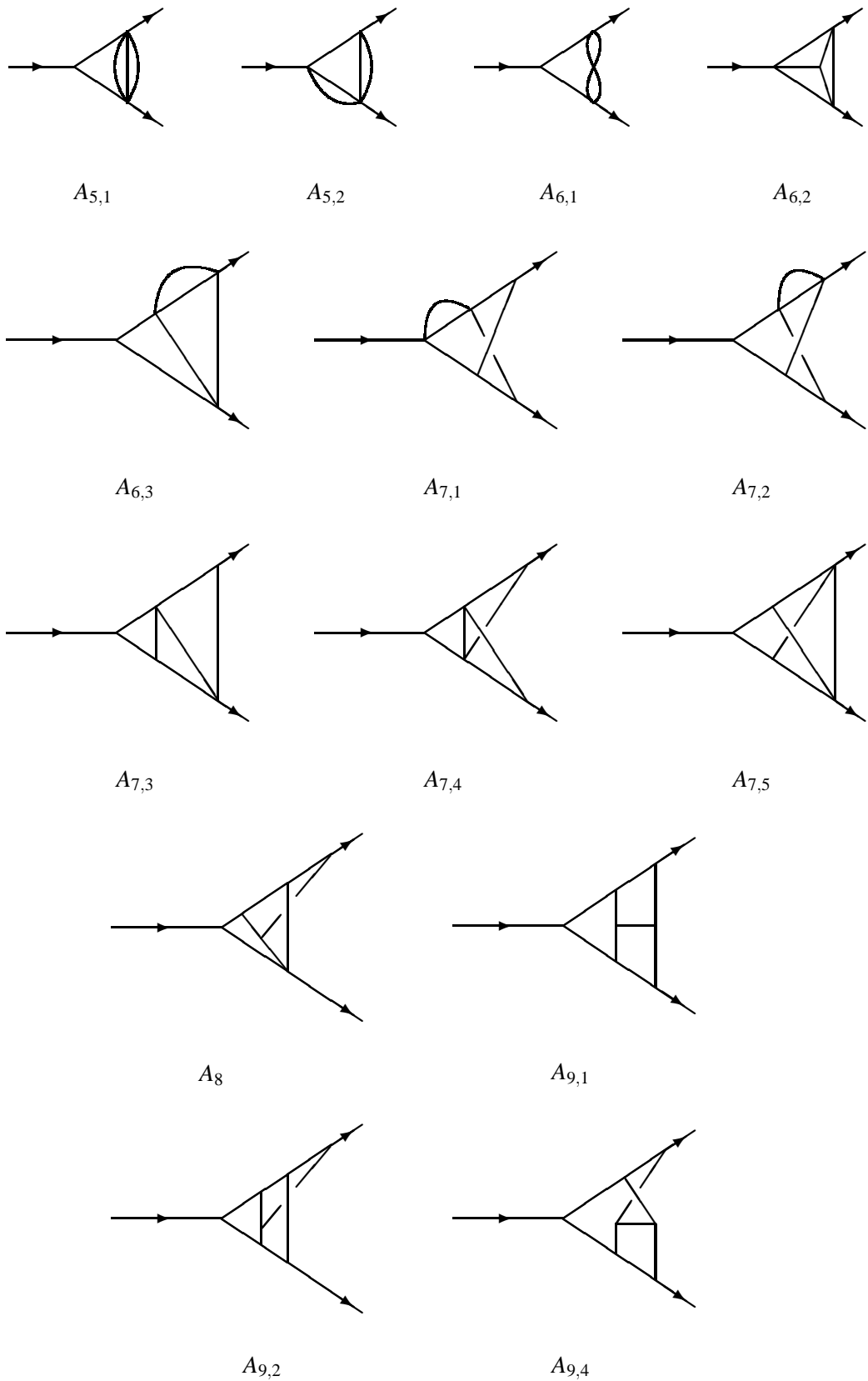

Below, we list the results~\cite{cedricpaper,Heinrich:2007at,Heinrich:2009be}
for the fourteen genuine vertex-type master integrals from
Fig.~\ref{fig:3loopdiagrams}. We use the
following definitions,
\begin{equation}
 q^2=(p_1+p_2)^2 \; , \qquad\quad \loopinttwo{dk} \equiv \loopint{k} \; , \qquad\quad S_{\Gamma} = \frac{1}{\lk
 4\pi\rk^{D/2}\,\Gamma(1-\e)} \; ,
\end{equation}
and moreover we tacitly assume that all propagators contain an infinitesimal $+i\eta$ with $\eta>0$.

\subsection{Integrals with five or six propagators}
We start with three integrals that can be displayed in a closed form in terms of $\Gamma$-functions only. Their expansion about
$\eps=0$ can be easily performed with standard computer algebra programs.
\bea
\dps A_{5,1} &=& \loopinttwo{dk}\loopinttwo{dl}\loopinttwo{dr} \;
\frac{1}{(k+p_1)^2 \; (k-p_2)^2 \; l^2 \; (k+l+r)^2\; r^2} \nnb \\
&&\nnb \\
&=& i \, \ESGamma^3 \, \lek -\lp q\rp^2-i \, \eta \rek^{1-3\, \eps} \frac{\Gamma^6(1-\eps) \, \Gamma(2\,\eps) \,
\Gamma(3\,\eps) \, \Gamma(1-3\,\eps)}{(1-2\,\eps) \, (2-3\,\eps) \, \Gamma(3-4\,\eps)} \; \; , \\
&&\nnb \\
%%%%%%%%%%%%%%%%%%%%%%%%%%%%%%%%%%%%%%%%%%%%%%%%%%%%%%%%%%%%%%%%%%%%%%%%%%%%%%%%%%%%%%%%%%%%%%%%%%%%%%%%%
\dps A_{5,2} % &=& \loopinttwo{dk}\loopinttwo{dl}\loopinttwo{dr} \;
%\frac{1}{(k+p_1)^2\; (l-k+p_2)^2\; l^2 \; (k+r)^2 \; r^2} \nnb \\
%&&\nnb \\
&=& - i \, \ESGamma^3 \, \lek -\lp q\rp^2-i \, \eta \rek^{1-3\, \eps} \frac{\Gamma^7(1-\eps) \, \Gamma(\eps) \,
\Gamma(3\,\eps) \, \Gamma(1-3\,\eps)}{(1-2\,\eps) \, \Gamma(2-2\,\eps) \, \Gamma(3-4\,\eps)} \; \; , \\
&&\nnb \\
%%%%%%%%%%%%%%%%%%%%%%%%%%%%%%%%%%%%%%%%%%%%%%%%%%%%%%%%%%%%%%%%%%%%%%%%%%%%%%%%%%%%%%%%%%%%%%%%%%%%%%%%%
\dps A_{6,1} % &=& \loopinttwo{dk}\loopinttwo{dl}\loopinttwo{dr} \;
%\frac{1}{(k+p_1)^2 \; (k-p_2)^2 \; l^2 \; (l+k)^2 \; r^2 \; (r+k)^2}\nnb \\
%&&\nnb \\
&=& - i \, \ESGamma^3 \, \lek -\lp q\rp^2-i \, \eta \rek^{-3\, \eps} \frac{\Gamma^7(1-\eps) \, \Gamma^2(\eps) \,
\Gamma(3\,\eps) \, \Gamma^2(1-3\,\eps)}{\Gamma^2(2-2\,\eps) \, \Gamma(2-4\,\eps)} \; \; .
\eea
The next integral, $A_{6,2}$, can be written in terms of a two-fold Mellin-Barnes representation.
\bea
\dps A_{6,2} % &=& \loopinttwo{dk}\loopinttwo{dl}\loopinttwo{dr} \;
%\frac{1}{\lk k+p_1\rk^2 \; \lk k+l-p_2\rk^2 \; \lp l \rp^2 \; \lp r \rp^2 \; \lk r-k\rk^2 \; \lk r-k-l\rk^2} \nnb \\
&=& -i \, \ESGamma^3 \, \lek -\lp q\rp^2-i \, \eta \rek^{-3\, \eps} \frac{\Gamma^3(1-\eps) \,
\Gamma(3\,\eps) \, \Gamma^2(1-3\,\eps)}{\Gamma(1-2\,\eps) \, \Gamma(2-4\,\eps)} \MB{c_1}{w_1} \MB{c_2}{w_2} \nnb \\
&&\nnb \\
&& \times \,
\frac{\Gamma(-1+3\,\eps-w_1)\,\Gamma(-1+2\,\eps-w_1)\,\Gamma(2-4\,\eps+w_1)\,\Gamma(-w_2)\,\Gamma(w_2-w_1)}{\Gamma(3\,\eps-w_1) \,
\Gamma(2-4\,\eps+w_2) \, \Gamma(2-4\,\eps+w_1-w_2)}\nnb \\
&&\nnb \\
&& \times \, \Gamma(1-\eps+w_2)\,\Gamma(1-\eps+w_1-w_2)\,\Gamma(1-2\,\eps+w_2)\,\Gamma(1-2\,\eps+w_1-w_2) \;\; . \label{eq:A62}
\eea
The contour integrals in the complex plane can be chosen as straight lines parallel to the imaginary axis, 
\textit{i.e.}\ the real parts $c_1$ and $c_2$ along the curves are constant. Their values, together with that of $\eps$,
must be chosen such as to separate left poles of $\Gamma$-functions from right ones~\cite{Tausk,smirnovbook,Alejandro}, which is
achieved by taking $c_1 = - 6/5$, $c_2 = - 1/2$, and $ -1/15 < \eps < 3/20$. The analytic continuation to $\eps=0$ is therefore trivial,
and after summing the residues one obtains
\bea
\dps A_{6,2} &=& i \, \ESGamma^3 \lek -\lp q\rp^2-i \, \eta \rek^{-3\, \eps}
 \, \lek -\frac{2 \, \zeta_3}{\eps} - 18 \, \zeta_3 -
\frac{7\pi^4}{180} + \lk  - 122 \, \zeta_3 - \frac{7\pi^4}{20} + \frac{2\pi^2}{3} \, \zeta_3 - 10 \, \zeta_5\rk \, \eps\rp\nnb \\
&&\nnb \\
&& \lp + \lk -738 \, \zeta_3 - \frac{427\pi^4}{180} + 6 \pi^2 \, \zeta_3 - 90 \, \zeta_5+ \frac{163\pi^6}{7560} + 76 \,
\zeta_3^2\rk \, \eps^2 + {\cal O} (\eps^3)\rek \; \; .
\eea
In Ref.~\cite{Huber:2007dx}, two more orders of the $\e$-expansion can be found. The next integral, $A_{6,3}$, can be displayed in a
closed form valid to all orders in $\eps$ in terms of hypergeometric functions of unit argument.
The latter are expanded in $\eps$ by means of the {\tt Mathematica} package {\tt
HypExp}~\cite{hypexp,Huber:2007dx},
\bea
\dps A_{6,3} % &=& \loopinttwo{dk}\loopinttwo{dl}\loopinttwo{dr} \; \; \frac{1}{\lp k\rp^2 \; \lk k-q\rk^2 \; \lk k-l\rk^2  \;
%\lk l-p_1\rk^2 \; \lk r-l\rk^2\; \lp r \rp^2} \nnb \\
%&&\nnb \\
&=& -i \, \ESGamma^3 \, \, \lek -\lp q\rp^2-i \, \eta \rek^{-3\, \eps} \, 
\frac{\Gamma^7(1-\eps) \, \Gamma(-1+3\,\eps)}{(1-2\,\eps)^2 \, (1-3\,\eps) \, \Gamma(2-4\,\eps)} \nnb \\
&&\nnb \\
&& \times \bigg[ -\frac{\Gamma(2-3\,\eps) \, \Gamma(2\,\eps) \, \Gamma(\eps) \,
\Gamma(2-2\,\eps)}{\Gamma^2(1-\eps)} \, + \,
\pFq{3}{2}{1,1-\eps,1-2\,\eps}{2-2\,\eps,2-3\,\eps}{1}\bigg] \nnb\\
&&\nnb\\
&=& i \, \ESGamma^3 \lek -\lp q\rp^2-i \, \eta \rek^{-3\, \eps}
\lek  - \frac{1}{6\,{\eps}^3} - \frac{3}{2\,{\eps }^2} - \lk
   \frac{55}{6} + \frac{{\pi }^2}{6}\rk\frac{1}{\eps } -\frac{95}{2}  - \frac{3\,{\pi }^2}{2}+ 
  \frac{17\,\zeta_3}{3} + \lk - \frac{1351}{6}  \rp \rp\nnb \\
&&\nnb \\
&& \lp - \frac{55\,{\pi }^2}{6}- \frac{{\pi }^4}{90} + 
     51\,\zeta_3 \rk \, \eps  + \lk - \frac{2023}{2}  - 
     \frac{95\,{\pi }^2}{2} - \frac{{\pi }^4}{10} + 
     \frac{935\,\zeta_3}{3} + 
     \frac{10\,{\pi }^2\,\zeta_3}{3} + 
     65\,\zeta_5 \rk \, \eps^2\, \nnb \\
&&\nnb \\
&& + \! \lk \! \frac{7\,{\pi }^6}{54}- \frac{26335}{6}   - 
     \frac{1351\,{\pi }^2}{6} - \frac{11\,{\pi }^4}{18} \lp  + 1615\,\zeta_3 + 
     30\,{\pi }^2\,\zeta_3 - 
     \frac{268\,\zeta_3^2}{3} + 585\,\zeta_5 \!\!
     \rk \! \eps^3 \! + {\cal O} (\eps^4)\rek \!\! . \nnb \\
\eea
\subsection{Integrals with seven or eight propagators}
Since the explicit expressions for the integrals that follow are very lengthy, we summarize in this paragraph the respective techniques
and subsequently only give the results of the Laurent expansion in $\eps$. Integrals $A_{7,1}$, $A_{7,2}$, and $A_{7,4}$ reveal a closed
form in terms of hypergeometric functions of unit argument, whose $\eps$ expansions are carried out with {\tt
HypExp}~\cite{hypexp,Huber:2007dx}. Integrals $A_{7,3}$, $A_{7,5}$, and $A_{8}$ possess multiple Mellin-Barnes representations which are
three-, four- and four-fold respectively. Only in the case of $A_8$ the analytic continuation to $\eps=0$ is non-trivial and is carried
out with {\tt MB}~\cite{czakonMB}. The explicit results read
\bea
\dps A_{7,1} % &=& \loopinttwo{dk}\loopinttwo{dl}\loopinttwo{dr} \; \; \frac{1}{\lp r\rp^2 \, \lk r-k\rk^2 \, \lk k-q\rk^2  \,
%\lk k-l\rk^2 \, \lk k-l-p_2\rk^2\, \lp l \rp^2 \, \lk l-p_1\rk^2} \nnb \\
%&&\nnb \\
&=& i \, \ESGamma^3 \lek -\lp q\rp^2-i \, \eta \rek^{-1-3\, \eps}
\lek  \frac{1}{4\,\eps^5} + \frac{1}{2\,\eps^4} + \lk 1 - \frac{\pi^2}{6}\rk\frac{1}{\eps^3}
+ \lk 2 - \frac{\pi^2}{3} - 10\,\zeta_3\rk\frac{1}{\eps^2} \rp\nnb \\
&&\nnb \\
&& + \lk 4 - \frac{2\,{\pi }^2}{3} - \frac{11\,{\pi }^4}{45} - 20\,\zeta_3\rk \, \frac{1}{\eps} 
 + \lk 8 - \frac{4\,{\pi }^2}{3}
- \frac{22\,{\pi }^4}{45} - 40\,\zeta_3 + \frac{14\,{\pi }^2\,\zeta_3}{3} - 88\,\zeta_5 \rk  \nnb \\
&&\nnb \\
&& + \lk 16 - \frac{8\,{\pi }^2}{3} - \frac{44\,{\pi }^4}{45} - 
    \frac{943\,{\pi }^6}{7560} - 80\,\zeta_3 \lp + \frac{28\,{\pi }^2\,\zeta_3}{3} + 196\,\zeta_3^2 - 176\,\zeta_5  \rk \eps + {\cal O}
    (\eps^2)\rek \; , \nnb \\
&& \\
\dps A_{7,2} % &=& \loopinttwo{dk}\loopinttwo{dl}\loopinttwo{dr} \; \frac{1}{\lp k\rp^2 \, \lk k-q\rk^2  \,
%\lk l-p_1\rk^2 \,\lk k-l\rk^2 \, \lk k-l-p_2\rk^2\, \lp r \rp^2 \, \lk r-l\rk^2} \nnb \\
%&&\nnb \\
&=& i \, \ESGamma^3 \lek -\lp q\rp^2-i \, \eta \rek^{-1-3\, \eps}
\lek  \frac{\pi^2}{12\,\eps^3} + \lk \frac{\pi^2}{6} + 2 \,\zeta_3 \rk\frac{1}{\eps^2}
+ \lk \frac{\pi^2}{3}+\frac{83 \,\pi^4}{720}+4\,\zeta_3\rk\frac{1}{\eps} \rp\nnb \\
&&\nnb \\
&& \hs{18} + \lk  \frac{2\,\pi^2}{3}
+ \frac{83\,\pi^4}{360} + 8\,\zeta_3 - \frac{5\,\pi^2\,\zeta_3}{3} + 15\,\zeta_5 \rk  \nnb \\
&&\nnb \\
&& \hs{18} + \lk \frac{4\,\pi^2}{3} + \frac{83\,\pi^4}{180} + 
    \frac{2741\,\pi^6}{90720} + 16\,\zeta_3 \lp - \frac{10\,\pi^2\,\zeta_3}{3} - 73\,\zeta_3^2 + 30\,\zeta_5  \rk \eps + {\cal O} (\eps^2)\rek \; , \\
    &&\nnb \\
\dps A_{7,3} % &=& \loopinttwo{dk}\loopinttwo{dl}\loopinttwo{dr} \; \frac{1}{\lp k\rp^2 \, \lk k+q\rk^2  \,
%\lk l-k-p_2\rk^2 \,\lk l-p_2\rk^2 \, \lk r+l\rk^2\, \lp r \rp^2 \, \lk r-p_1\rk^2} \nnb \\
%&&\nnb\\
&=& i \, \ESGamma^3 \lek -\lp q\rp^2-i \, \eta \rek^{-1-3\, \eps} \,\lek \lk -\frac{\pi^2\,\zeta_3}{6}-10
\,\zeta_5\rk\frac{1}{\eps} - \frac{119 \,\pi^6}{2160} - \frac{31}{2} \, \zeta_3^2 + {\cal O} (\eps)\rek , \\
&& \nnb\\
\dps A_{7,4} % &=& \loopinttwo{dk}\loopinttwo{dl}\loopinttwo{dr} \; \frac{1}{\lp k\rp^2 \, \lk k-q\rk^2  \,
%\lk r+l-k\rk^2 \,\lp l\rp^2 \, \lk l-p_1\rk^2\, \lp r \rp^2 \, \lk r-p_2\rk^2} \nnb \\
%&&\nnb\\
&=& i \, \ESGamma^3 \, \lek -\lp q\rp^2-i \, \eta \rek^{-1-3\, \eps}
 \bigg[ \frac{6\,\zeta_3}{\eps ^2} + \lk \frac{11\,\pi^4}{90} + 36\,\zeta_3 \rk\frac{1}{\eps} + 
 \lk\frac{11\,{\pi }^4}{15} + 216\,\zeta_3 - 2\,{\pi }^2\,\zeta_3 + 46\,\zeta_5\rk \nnb \\
&&\nnb \\
&&\hs{15}  + \lk \frac{22\,{\pi }^4}{5} - \frac{19\,{\pi }^6}{270} +  1296\,\zeta_3 - 12\,{\pi }^2\,\zeta_3 - 
    282\,\zeta_3^2 + 276\,\zeta_5 \rk \, \eps  + {\cal O}(\eps^2) \bigg] ,  \\
&&\nnb \\
\dps A_{7,5} % &=& \loopinttwo{dk}\loopinttwo{dl}\loopinttwo{dr} \; \frac{1}{\lp k\rp^2 \, \lk k+q\rk^2  \,
%\lk k+r\rk^2 \,\lk l-p_2\rk^2 \, \lk r-l\rk^2\, \lp r \rp^2 \, \lk k+l+p_1\rk^2} \nnb \\
%&&\nnb\\
&=& i \, \ESGamma^3 \lek -\lp q\rp^2-i \, \eta \rek^{-1-3\, \eps} \,\lek  2\pi^2\,\zeta_3+10
\,\zeta_5 + \lk 12\pi^2\,\zeta_3+60
\,\zeta_5+\frac{11 \pi^6}{162} + 18 \, \zeta_3^2\rk \eps + {\cal O} (\eps^2)\rek  . \nnb \\
&&\\
\dps A_8 % &=& \loopinttwo{dk}\loopinttwo{dl}\loopinttwo{dr} \; \frac{1}{\lk k+p_1\rk^2 \, \lk k+r\rk^2  \,
%\lk k+r+q\rk^2 \,\lk l-k\rk^2 \, \lk l+r\rk^2\, \lp l \rp^2 \, \lp r \rp^2 \, \lk l+p_1\rk^2} \nnb \\
%&&\nnb\\
&=& i \, \ESGamma^3 \lek -\lp q\rp^2-i \, \eta \rek^{-2-3\, \eps} \, \bigg[ \frac{8\zeta_3}{3\eps^2} + \lk\frac{5\pi^4}{27}-8
\zeta_3\rk\frac{1}{\eps} + 24\zeta_3-\frac{5\pi^4}{9}-\frac{52}{9} \, \pi^2\zeta_3 +\frac{352}{3} \,\zeta_5 \nnb \\
&&\nnb \\
&&\hs{15}  + \lk - 72\zeta_3+\frac{5\pi^4}{3}+\frac{52}{3} \, \pi^2\zeta_3 -352 \,\zeta_5
+\frac{1709\pi^6}{8505}-\frac{332}{3}\zeta_3^2\rk \, \eps  + {\cal O}(\eps^2) \bigg] .  \label{eq:A8erg}
\eea
\subsection{Integrals with nine propagators}
Integrals $A_{9,1}$, $A_{9,2}$, and $A_{9,4}$ possess a six-fold MB representation each. The analytic continuation, carried out with
{\tt MB}~\cite{czakonMB}, involves in each case approximately 200 steps. We apply the techniques described in
section~\ref{sec:compmethods}, and stress that all results were obtained by purely analytic steps. We have
\bea
\dps A_{9,1} &=& \loopinttwo{dk}\loopinttwo{dl}\loopinttwo{dr} \; \frac{1}{\lp k\rp^2 \, \lk k+p_1\rk^2 \, \lk k+l\rk^2  \,
\lk k-r\rk^2 \,\lk l+r\rk^2 \, \lk l+p_2\rk^2\, \lp l \rp^2 \, \lk r+p_1 \rk^2 \, \lk r-p_2\rk^2} \nnb \\
&&\nnb\\
&=& i \, \ESGamma^3 \, \lek -\lp
q\rp^2-i \, \eta \rek^{-3-3\, \eps} \Big[-\frac{1}{18
    \e^5}+\frac{1}{2 \e^4}+\Big(-\frac{53}{18}-\frac{4 \pi^2}{27}\Big)
  \frac{1}{\e^3}+\Big(\frac{29}{2}+\frac{22
    \pi^2}{27}-2\zeta_3\Big)\frac{1}{\e^2}\nnb\\
    && +\Big(\frac{158}{9}\zeta_3-\frac{129}{2}-\frac{8\pi^2}{3}-\frac{20\pi^4}{81}\Big)\frac{1}{\e}
     +\Big(\frac{537}{2}+6\pi^2-\frac{578}{9}\zeta_3+\frac{322\pi^4}{405}-\frac{14}{3}\pi^2\zeta_3-\frac{238}{3}\zeta_5\Big)\nnb\\
    && +\Big(-\frac{2133}{2}-4\pi^2+158\zeta_3-\frac{302\pi^4}{135}-\frac{26}{3}\pi^2\zeta_3+
    \frac{826}{3}\zeta_5-\frac{2398\pi^6}{5103}-\frac{466}{3}\zeta_3^2\Big)\e+{\cal{O}}(\e^2)\Big] \, , \nnb\\ \label{eq:A91expand}
\eea
\bea
\dps A_{9,2} &=& \!\loopinttwo{dk}\!\!\loopinttwo{dl}\!\!\loopinttwo{dr} \frac{1}{\lp k\rp^2 \! \lk k+p_1\rk^2 \! \lk k-l+p_1\rk^2  \!
\lk k-r-l\rk^2 \! \lk l+r\rk^2 \! \lk l+p_2\rk^2 \!\! \lp l \rp^2 \! \lk r+p_1 \rk^2 \! \lk r-p_2\rk^2} \nnb \\
&&\nnb\\
&=& i \, \ESGamma^3 \, \lek -\lp q\rp^2-i \, \eta \rek^{-3-3\, \eps}
\Big[\frac{2}{9 \e^6}+\frac{5}{6 \e^5}+\Big(-\frac{20}{9}-\frac{7 \pi^2}{27}\Big) \frac{1}{\e^4}+\Big(\frac{50}{9}-\frac{17
\pi^2}{27}-\frac{91}{9}\zeta_3\Big)\frac{1}{\e^3}\nnb\\
&& +\Big(-\frac{110}{9}+\frac{4\pi^2}{3}-\frac{166}{9}\zeta_3-\frac{373\pi^4}{1080}\Big)\frac{1}{\e^2}
+\Big(\frac{170}{9}-\frac{16\pi^2}{9}+\frac{494}{9}\zeta_3-\frac{187\pi^4}{540}\nnb\\
&& +\frac{179}{27}\pi^2\zeta_3-167\zeta_5\Big)\frac{1}{\e} +\left(-670.0785 \pm 0.0326 \right)+{\cal{O}}(\e)\Big] \; .\label{eq:A92expand}
\eea
\bea
\dps A_{9,4} &=& \loopinttwo{dk}\loopinttwo{dl}\loopinttwo{dr} 
\; \frac{1}{\lp k\rp^2 \lk k+p_1\rk^2 \lk k-r\rk^2 
\lk k-r-l\rk^2 \lk l+r\rk^2 \lk l+p_2\rk^2 \lp l \rp^2 \lk r+p_1 \rk^2 \lk r-p_2\rk^2} \nnb \\
&&\nnb\\
&=& i \, \ESGamma^3 \, \lek -\lp q\rp^2-i \, \eta \rek^{-3-3\, \eps}
\Big[\frac{1}{9 \e^6}+\frac{8}{9 \e^5}+\Big(-1-\frac{10 \pi^2}{27}\Big) \frac{1}{\e^4}+\Big(-\frac{14}{9}-\frac{47
\pi^2}{27}-12\zeta_3\Big)\frac{1}{\e^3}\nnb\\
&& +\Big(17+\frac{71\pi^2}{27}-\frac{200}{3}\zeta_3-\frac{47\pi^4}{810}\Big)\frac{1}{\e^2}+\Big(117.3999538\pm
0.0000032\Big)\frac{1}{\e}\nnb\\
&& +\left(1948.167043 \pm 0.000025 \right)+{\cal{O}}(\e)\Big] \; .\label{eq:A94expand}
\eea
The numbers were obtained with {\tt MB.m}~\cite{czakonMB}. We stress that all other terms
in~(\ref{eq:A91expand})~--~(\ref{eq:A94expand}) were derived by purely analytic steps. The analytic result
of the simple pole of $A_{9,4}$ can be extracted from~\cite{BCSSS}.
It turns out that for each of the above integrals a corresponding one with an irreducible scalar product in the
numerator can be chosen whose coefficients of the $\eps$-expansion have homogeneous weight. The suitably chosen numerators are
$r^2$ for $A_{9,1}$ and $A_{9,4}$, and $(l-p_1)^2$ for $A_{9,2}$. The property of homogeneous weight is very helpful when one uses the
PSLQ algorithm~\cite{pslq}, by means of which we obtain
\bea
\dps A_{9,1}^{(n)}&=& i \, \ESGamma^3 \, \lek -\lp q\rp^2-i \, \eta
\rek^{-2-3\, \eps} \Big[-\frac{1}{36
    \e^6}-\frac{\pi^2}{18\e^4}-\frac{14\zeta_3}{9\e^3}-\frac{47\pi^4}{405\e^2}\nnb\\ &&
  +\left(-\frac{85}{27}\pi^2\zeta_3-20\zeta_5\right)\frac{1}{\e}
  +\left(-\frac{1160\pi^6}{5103}-\frac{137}{3}\zeta_3^2\right)+{\cal{O}}(\e)\Big]
\; ,  \label{eq:A91nResult} \\
&&\nnb \\
\dps A_{9,2}^{(n)}&=& i \, \ESGamma^3 \, \lek -\lp q\rp^2-i \, \eta \rek^{-2-3\, \eps}
\Big[-\frac{2}{9 \e^6}+\frac{7\pi^2}{27\e^4}+\frac{91\zeta_3}{9\e^3}+\frac{373\pi^4}{1080\e^2}\nnb\\
&& +\left(-\frac{179}{27}\pi^2\zeta_3+167\zeta_5\right)\frac{1}{\e}
+\left(395.3405   \pm 0.0326  \right)+{\cal{O}}(\e)\Big] \; , \\
&&\nnb \\
\dps A_{9,4}^{(n)}&=& i \, \ESGamma^3 \, \lek -\lp q\rp^2-i \, \eta \rek^{-2-3\, \eps}
\Big[\frac{1}{9 \e^6}-\frac{10\pi^2}{27\e^4}-\frac{12\zeta_3}{\e^3}-\frac{47\pi^4}{810\e^2}\nnb\\
&& +\left(206.7612077   \pm 0.0000032  \right)\frac{1}{\e}
+\left( 1237.300592 \pm 0.000035  \right)+{\cal{O}}(\e)\Big] \; .
\eea
Since each topology contains only one master integral, each $A_{9,i}$ is related to its corresponding $A^{(n)}_{9,i}$.
We established these relations with the Laporta algorithm~\cite{laporta,air,fire} and checked that they are filfilled by the above
expressions.

The analytic expressions for the remaining coefficients are within reach. In the finite part of $A_{9,2}$ we are missing only $\sim 30$
terms, all four- and five-fold MB integrals. From what we can judge it is not possible to process these MB kernels by purely analytic
steps. We therefore plan to break them down to lower-dimensional integrals over ordinary Feynman parameters
and then use the PSLQ algorithm. In the case of $A_{9,4}$ we are left with ${\cal O}(10^3)$ MB terms which are at most three-fold. We
are confident that they can be processed by purely analytic steps.

\section*{Acknowledgements}
I would like to thank the organizers of RADCOR~2009 for creating a pleasant and inspiring atmosphere.
Special thanks goes to my colleagues from~\cite{cedricpaper,Heinrich:2007at,Heinrich:2009be} for a fruitful collaboration. The work
of the author was supported by SFB/TR~9 and by the Helmholtz-Alliance ``Physics at the Terascale''.

\end{document}